\documentclass{article}

\usepackage{amssymb}

\newtheorem{theorem}{Theorem}

\newtheorem{lemma}[theorem]{Lemma}
\newtheorem{definition}[theorem]{Definition}

\newenvironment{proof}{\noindent\bf{Proof.}\rm}{\hfill$\blacksquare$\bigskip}

\begin{document}

\title{A randomized strategy in the mirror game}

\author{Uriel Feige~\thanks{Department of Computer Science and Applied Mathematics, The Weizmann Institute. {\tt uriel.feige@weizmann.ac.il}}}

\maketitle

\begin{abstract}
Alice and Bob take turns (with Alice playing first) in declaring numbers from the set $[1,2N]$. If a player declares a number that was previously declared, that player looses and the other player wins. If all numbers are declared without repetition, the outcome is a tie. If both players have unbounded memory and play optimally, then the game will be tied. Garg and Schneider [ITCS 2019] showed that if Alice has unbounded memory, then Bob can secure a tie with $\log N$ memory, whereas if Bob has unbounded memory, then Alice needs memory linear in $N$ in order to secure a tie.

Garg and Schneider also considered an {\em auxiliary matching} model in which Alice gets as an additional input a random matching $M$ over the numbers $[1,2N]$, and storing this input does not count towards the memory used by Alice. They showed that is this model there is a strategy for Alice that ties with probability at least $1 - \frac{1}{N}$, and uses only $O(\sqrt{N} (\log N)^2)$ memory. We show how to modify Alice's strategy so that it uses only $O((\log N)^3)$ space.
\end{abstract}

\section{Introduction}

Garg and Schneider~\cite{GS} introduced a game between two players, Alice and Bob, which they referred to as the {\em mirror game}. Alice and Bob take turns (with Alice playing first) in declaring numbers from the set $[1,2N]$. If a player declares a number that was previously declared, that player looses and the other player wins. If all numbers are declared without repetition, the outcome is a tie. (The outcomes that we consider to be a tie are considered in \cite{GS} to be a win for both players, but this difference is irrelevant to this manuscript.) If both players have unbounded memory and play optimally, then the game will be tied. Garg and Schneider studied the situation in which one player has bounded memory whereas the other player has unbounded memory.

If Alice is the player with unbounded memory, then Bob can secure a tie with $\log N$ memory (just remembering the latest number declared by Alice). This can be done via the following {\em mirror strategy}. Fix a matching $M$ between numbers (e.g., number $x$ is matched with $2N+1-x$). We refer to two numbers that are matched to each other under $M$ as {\em partners} of each other. At every even round (these are the rounds in which Bob speaks), Bob declares the partner of the number declared by Alice in the previous round.

If Bob is the player with unbounded memory, then it is proved in~\cite{GS} that Alice needs memory linear in $N$ in order to secure a tie. However, it is left open whether Alice has a randomized strategy that uses sublinear space and secures a tie with high probability (e.g., with probability at least $\frac{2}{3}$).

As a step towards addressing the open question, Garg and Schneider considered a model in which Alice gets as an additional input a matching $M$ over the numbers $[1,2N]$, and storing this input (in Read Only Memory) does not count towards the memory used by Alice. The matching $M$ is chosen uniformly at random, and is not known to Bob. It is shown that in this {\em auxiliary random matching} model, there is a strategy for Alice that uses only $O(\sqrt{N} (\log N)^2)$ memory and ties with probability at least $1 - \frac{1}{N}$. Garg and Schneider ask whether this upper bound of $\tilde{O}(\sqrt{N})$ is tight in this model (see open problem~3 in~\cite{GS}). We show that it is far from being tight. Specifically:

\begin{theorem}
\label{thm:main}
In the mirror game with auxiliary random matching, Alice has a strategy that ties with probability at least $1 - \frac{1}{N}$, and uses only $O((\log N)^3)$ memory.
\end{theorem}

\section{Proof of the main theorem}

We shall use a straightforward variation of a result that is also used in~\cite{GS}.

\begin{definition}
\label{def:MNP}
The $(N,s,k)$ {\em missing numbers problem} is defined as follows. For integers $1 \le k < s \le 2N$, the {\em offline input} is a set $S \subset [1,2N]$ of cardinality $|S| = s$. The {\em online input} is a set $S' \subset S$ of cardinality $s - k \le |S'| \le s$. The goal is to output the set $S \setminus S'$.
\end{definition}

\begin{lemma}
\label{lem:MNP}
There is a deterministic algorithm, with unlimited access to the offline input and streaming access (one pass) to the online input, that solves the $(N,s,k)$ missing numbers problem, and uses $O(k \log N)$ space.
\end{lemma}

The proof of Lemma~\ref{lem:MNP} for the case $s = N$ appears in~\cite{GS} and in~\cite{M}. That proof can be adapted also to the case $s < N$. We present a sketch of proof for completeness.

\begin{proof}
Choose an arbitrary prime $p$ satisfying $2N < p < 4N$. For every $1 \le i \le k$ compute $S_i = \sum_{x\in S} x^i$ modulo $p$ and $S'_i = \sum_{y\in S'} y^i$ modulo $p$. Such computations can be done using only one pass over $S'$, and with space $O(k\log N)$ (because there are $2k$ sums to compute in parallel, and each one is smaller than $p$). In addition, compute $|S'|$ and $k' = |S|-|S'|$ (namely, $k'$ is the cardinality of the set of missing numbers, $k' \le k$). Among all possible subsets of size $k'$ of $S$ (there are ${|S|\choose k'} \le 2^{O(k\log n)}$ such subsets to try), the set $T$ of missing numbers is precisely the one that satisfies $\sum_{z\in T} z^i = S_i - S'_i$ modulo $p$, for all $1 \le i \le k'$. Hence $T$ can be found using exhaustive search in space $O(k\log N)$. (The time efficiency of the algorithm can be substantially improved, but this is not a concern of ours in this manuscript.)
\end{proof}

Before presenting our proof for Theorem~\ref{thm:main}, let us discuss how the random matching $M$ is represented, as this affects our proof. So as to clarify that the representation of $M$ is not counted as part of the space used by Alice's algorithm, we consider a model in which Alice has oracle accesses to $M$. That is, the matching $M$ is held by some oracle, and Alice can query the oracle in order to obtain matched pairs of numbers. Any space that Alice uses in order to store the answers to her queries is counted as part of her work space. There are two natural query models that come to (the author's) mind.

\begin{description}

\item[Random matching.] The oracle holds one of $\frac{(2N)!}{N!2^N}$ possible matchings, chosen at random. A query $q$ is a number $q \in [1,2N]$ , and the answer to the query is the number $M(q)$ that is matched with $q$ under the random matching $M$.

\item[Random list.] The oracle holds one of $\frac{(2N)!}{2^N}$ ordered lists of matched pairs of numbers, chosen at random. The random list contains $N$ matched pairs of numbers, $(M_1, M_2, \ldots, M_N)$, giving the random matching $M$. A query $q$ is a number $q \in [1,N]$, and the answer to the query is the pair  $M_q$ of two matched numbers under the random matching $M$. In this query model, the answer to $q$ depends both on the matching $M$ and on the order among the matched pairs. Both the matching and the order are random.

\end{description}

Queries in the random list model can simulate queries in the random matching model (to find which number is matched to a number $x$, query the pairs one by one until the reply is a pair that contains $x$). However, queries in the random matching model cannot simulate queries in the random list model, as answering the random list queries consistently is equivalent to storing a random permutation on $N$ items (pairs of matched numbers), and Alice does not have sufficient space for this.

The proof that we present for Theorem~\ref{thm:main} assumes the random list model. This will allow Alice to use Lemma~\ref{lem:MNP} with sets $S$ that are random and not known to Bob. See further discussion in Section~\ref{sec:discuss}.

We now prove Theorem~\ref{thm:main}.

\begin{proof}
For simplicity of the notation, we assume that $N$ is of the form $N = 2^n$ for some integer $n$. (The proof can easily be adapted to hold without this assumption.) Let $M_1, \ldots, M_N$ be the pairs of matched numbers under the auxiliary matching $M$ (in the random list model). Partition the set $[1,2N]$ into $n$ sets, where set $P_1$ contains the four number in the pairs $M_1$ and $M_2$, and for $2 \le i \le n$ the set $P_i$ contains all those numbers that appear in the pairs $M_j$ with $2^{i-1} + 1 \le j \le 2^i$. Hence $P_i$ (for $i \ge 2$) has $2^{i-1}$ pairs of matched numbers, and $2^i$ numbers altogether. Fix $k = c\log N$, where $c$ is a sufficiently large constant ($c = 2$ suffices).

For every $i$ that satisfies $2^i \le k$, Alice's algorithm keeps track of all declared numbers within $P_i$. For every $i$ that satisfies $2^i > k$, the algorithm runs the $(N,s,k)$ missing number algorithm of Lemma~\ref{lem:MNP} on the declared numbers, with $S = P_i$ and $s = 2^i$. We say that $P_i$ is {\em exhausted} (at a certain time step) if all numbers in $P_i$ have already been declared. We say that $P_i$ is {\em explicit} (at a certain time step) if at most $k$ numbers in $P_i$ have not yet been declared. Note that for explicit $P_i$'s all their non-declared numbers are known to the Alice's algorithm, by Lemma~\ref{lem:MNP}.

Initially, Alice declares the first number in the pair $M_1$ (this number is in $P_1$). Call this number $g_1$ ($g$ stands for ``generate"). Thereafter, for every number that Bob declares, Alice responds with its partner under $M$. This continues until Bob declares $g_1$'s partner. At this point Alice, who cannot declare $g_1$ again,  finds the smallest index $i \ge 1$ for which $P_i$ is not exhausted. If this $P_i$ is explicit then Alice picks the first yet undeclared number from $P_i$ (call it $g_2$), and declares it. (If $P_i$ is not explicit then Alice {\em aborts} and looses the game.) Now Alice continues with the mirror strategy until Bob declares $g_2$'s partner. Again, Alice finds the smallest index $i$ for which $P_i$ is not exhausted, and if $P_i$ is explicit then Alice picks the first yet undeclared number from $P_i$ (call it $g_3$), and declares it. This form of strategy is continued by Alice, and if all numbers are exhausted without Alice aborting, the game ends in a tie.

The algorithm can be implemented (in the random list model) using a space of $n$ (this is the number of sets $P_i$) times $O(k \log N)$ (from Lemma~\ref{lem:MNP}). As $k = c\log N$ and $n = \log N$, the total space is $O((\log N)^3)$, as desired.

It remains to upper bound the probability that Alice aborts. For $2 \le i \le n$, let $B_i$ denote the bad event that there is some time step during the game in which $P_i$ is not yet explicit whereas all of $P_1, \ldots, P_{i-1}$ are already exhausted. For Alice to abort, it must happen that at least one of the bad events $B_i$ happens. Hence let us fix one value $2 \le i \le n$ and upper bound the probability of event $B_i$. We use the following observations.

\begin{enumerate}

\item Initially, the number of matched pairs in $P_i$ equals the number of matched pairs in $\bigcup_{j=1}^{i-1} P_j$.

\item Any number declared by Bob that is neither in $P_i$ nor in $\bigcup_{j=1}^{i-1} P_j$ is irrelevant to the event $B_i$.

\item Whenever it is Bob's turn to declare a number, given that the number that Bob declares is from $\bigcup_{j=1}^{i} P_j$, the probability (taken over the randomness of the random list oracle) that this number is from $P_i$ exactly equals the fraction of yet undeclared numbers in $\bigcup_{j=1}^{i} P_j$ that are in $P_i$.

\item When Alice generates a number $g_{\ell}$ from $\bigcup_{j=1}^{i-1} P_j$ (following a declaration by Bob of the partner of $g_{\ell - 1}$), the partner of $g_{\ell}$ (under $M$) has not been declared yet. Hence the number of undeclared pairs in $\bigcup_{j=1}^{i-1} P_j$ does not change, whereas the number of numbers in $\bigcup_{j=1}^{i-1} P_j$ decreases by~1. This increases the likelihood that in future time steps Bob will declare a number from $P_i$ (and decreases the likelihood that  Bob will declare a number from $\bigcup_{j=1}^{i-1} P_j$).

\end{enumerate}

The event $B_i$ can thus be modelled by the following {\em Alice starts} two-bin process. There are two bins, $A$ (representing the numbers is $\bigcup_{j=1}^{i-1} P_j$) and $B$ (representing the numbers in $P_i$), each initially containing $2^{i+1}$ ``balls". First, Alice draws one of the balls out of bin $A$ (this is the number $g_1$). Thereafter, at each round, Bob draws a ball $b$, chosen uniformly at random among all remaining balls, and Alice draws another ball from the same bin as $b$. The event $B_i$ happens if bin $B$ contains more than $k$ balls at the time that bin $A$ becomes empty. Rather than analysing the {\em Alice starts} two-bin process, we consider a {\em Bob starts} two-bin process which differs from the above only in the fact that Bob is the first to draw a ball (rather than Alice starting with $g_1$). Using Observation~4 above, the probability of $B_i$ in the {\em Bob starts} process is at least as high as it is in the {\em Alice starts} process. The probability of $B_i$ in the {\em Bob starts} process is exactly the same as the probability that in a random permutation over $2^{i+1}$ items (where each item corresponds to a matched pair of numbers), half of which are in bin $A$ and half in bin $B$, the last $k+1$ items are all in $B$. This probability is smaller than $2^{-k}$.

A union bound over all $i$ shows that the probability of Alice aborting is at most $n2^{-k} = \log N \cdot N^{-c} \le \frac{1}{N}$ (when the constant $c$ in the definition of $k = c\log N$ is sufficiently large), as desired.
\end{proof}

\section{Discussion}
\label{sec:discuss}

Theorem~\ref{thm:main} is proved in the random list model. The given proof requires that both the matching $M$ and the sets $P_i$ are random, so as to be hidden from Bob. Throughout the rounds of the game, Bob gains much information about the auxiliary random matching $M$. Bob also gains some information about the sets $P_i$ (which are determined by the random order of the matched pairs, as held by the oracle in the list model), but this information is quite limited. Given the information that Bob learns about $M$, the additional information that Bob learns about the sets $P_i$ is not more (and even less) than whatever is implied by the sequence $g_1, g_2, \ldots $ of numbers generated by Alice (those that are not parters on numbers previously declared by Bob), and the time steps in which each such number is generated. As the expected length of this sequence is $O(\log N)$, Bob learns (in expectation) only $O(\log^2 N)$ bits of information about the sets $P_i$. For this reason, we believe that (with additional work) the proof of Theorem~\ref{thm:main} can also be extended to the random matching model, by having Alice generate the sets $P_i$ by herself in a pseudo-random way. However, no attempt is made in this manuscript to turn this intuitive argument into a formal proof.

Theorem~\ref{thm:main} that applies in the random list model easily extends to some related models that do not involve an oracle. Here are three such models:

\begin{description}

\item [Preprocessing model.] Before the game begins, Alice is allowed to write to herself some polynomial size advice on some auxiliary read only memory. After the game begins, she may consult the advice, but cannot add any new advice. In this model, Alice can write to herself as advice a random list of matched pairs, and Theorem~\ref{thm:main} applies.
    
\item [Auxiliary random string.] In this model, before the game begins, a polynomial length string of random bits is written into auxiliary memory. Alice has read only access to this memory and Bob has no access to it. If this string is of length $c N (\log N)^2$, where $c$ is a sufficiently large constant, then it can be interpreted by Alice in a natural way as a matching $M$, represented as in the random list model. Break the string into roughly, $cN\log N$ words, where each word contains $1 + \log N$ bits. The words naturally correspond to numbers in the range $[1,2N]$, and with high probability, each such number appears at least once in the list. This induces a random permutation over the numbers $[1,2N]$, according to the first appearance of each number. For every $q$, if Alice wishes to find the $q$th number in the permutation, she can do so using $O(\log N)$ work space, using repeated scans of the random string (details omitted). Consequently, Theorem~\ref{thm:main} applies in this model.

\item [Randomized Bob.] Bob uses the {\em uniformly random} strategy. In this strategy, whenever it is Bob's turn to declare a number, he chooses a number uniformly at random among those numbers not yet declared. In this case the matching $M$ can be arbitrary and need not be given to Alice as a separate input. For example, one can fix $M$ to be the matching in which $M_i = (i, 2N + 1 - i)$. Theorem~\ref{thm:main} implies that Alice can tie with probability at least $1 - \frac{1}{N}$ using a deterministic strategy that uses only $O((\log N)^3)$ memory. Here, probability is computed over the random choices of Bob.

\end{description}

\subsection*{Acknowledgements}

The work of the author is supported in part by the Israel Science Foundation (grant No. 1388/16).

\end{document}